\documentclass[a4paper,12pt]{article}

\usepackage{graphics}
\usepackage{latexsym}
\title{\textbf{\LARGE{Tachyonic Kink and Lump-like Solutions \vspace{0.5cm}\\ 
in Superstring Field Theory }} \\ \vspace{1.2cm}}
\author{Kazuki Ohmori\thanks{E-mail: 
ohmori@hep-th.phys.s.u-tokyo.ac.jp}
\vspace{1cm} \\
\small{\textit{Department of Physics, Faculty of Science, University of 
Tokyo}} \\
\small{\textit{Hongo 7-3-1, Bunkyo-ku, Tokyo 113-0033, Japan}}}
\date{}
\newcommand{\ap}{\alpha^{\prime}}
\newcommand{\cA}{\mathcal{A}}
\newcommand{\cB}{\mathcal{B}}

\newcommand{\cE}{\mathcal{E}}
\newcommand{\cF}{\mathcal{F}}
\newcommand{\cG}{\mathcal{G}}
\newcommand{\cH}{\mathcal{H}}
\newcommand{\cI}{\mathcal{I}}
\newcommand{\cJ}{\mathcal{J}}

\newcommand{\cM}{\mathcal{M}}
\newcommand{\cN}{\mathcal{N}}

\newcommand{\cS}{\mathcal{S}}
\newcommand{\cT}{\mathcal{T}}

\newcommand{\zetto}{\mathbf{Z}}
\newcommand{\aaru}{\mathbf{R}}

\newcommand{\gh}{\#_{\mathrm{gh}}}
\newcommand{\pic}{\#_{\mathrm{pic}}}
\newcommand{\ca}{{}^{\circ}\!\!\!\>\!{}_{\circ}}
\newcommand{\llk}{\langle\!\langle}
\newcommand{\rrk}{\rangle\!\rangle}
\newcommand{\bllk}{\biggl\langle\!\!\!\biggl\langle}
\newcommand{\brrk}{\biggr\rangle\!\!\!\biggr\rangle}

\begin{document}
\begin{titlepage}
\thispagestyle{empty}
\begin{flushright}
UT-934 \\
hep-th/0104230 \\
April, 2001 
\end{flushright}

\vskip 1.5 cm

\begin{center}
\noindent{\textbf{\LARGE{Tachyonic Kink and Lump-like Solutions \vspace{0.5cm}\\ 
in Superstring Field Theory }}}
\vskip 1.5cm
\noindent{\large{Kazuki Ohmori}\footnote{E-mail: 
ohmori@hep-th.phys.s.u-tokyo.ac.jp}}\\ 
\vspace{1cm}
\noindent{\small{\textit{Department of Physics, Faculty of Science, University of 
Tokyo}} \\ \vspace{2mm}
\small{\textit{Hongo 7-3-1, Bunkyo-ku, Tokyo 113-0033, Japan}}}
\end{center}
%\maketitle
\vspace{1cm}
\begin{abstract}
We construct a kink solution on a non-BPS D-brane using Berkovits' formulation of 
superstring field theory in the level truncation scheme. 
The tension of the kink reproduces 95\% of the expected 
BPS D-brane tension. We also find a lump-like solution which is interpreted as a kink--antikink 
pair, and investigate some of its properties. These results may be considered as successful tests of 
Berkovits' superstring field theory combined 
with the modified level truncation scheme. 
\end{abstract}
\end{titlepage}
\newpage
%\tableofcontents
\baselineskip 6mm

%%%%%main text

\section{Introduction}
In a couple of years, Sen's conjectures on the tachyon condensation 
have successfully been examined using various types of 
open string field theories.\footnote{For a review, see \cite{KO}.} In Witten's cubic string field 
theory (CSFT)\cite{csft}, not only have the (approximate) tachyon potential~\cite{potential,SZ} and 
lower dimensional D-branes as tachyonic lump solutions been constructed~\cite{lump,MSZ}, but also the 
absence of physical open string excitations (more precisely, the triviality of the cohomology 
classes) at the minimum of the tachyon potential has recently been verified in~\cite{noopen}. 
In boundary string field theory (BSFT)~\cite{bsft}, 
the exact tachyon potential and lower dimensional 
D-branes have been worked out both in bosonic string theory~\cite{KMM1} and in superstring 
theory (\cite{KMM2} for non-BPS D-branes and \cite{TTU} for the brane-antibrane system). 
However, only a limited amount of works have been done using Berkovits' Wess-Zumino-Witten--like 
open superstring field theory~\cite{Berkovits}. So far, the universal tachyon 
potential~\cite{NS,BSZ,3220,IqNaq1}, the effective potential for marginal deformation parameter 
of string field theory~\cite{IqNaq2} and the tachyon potential in the D0/D4 system with a 
Neveu-Schwarz $B$-field~\cite{David} have been obtained. 
One of our aims of this paper is to provide this theory 
with a new piece of evidence that it correctly describes 
the dynamics of open superstrings by constructing 
tachyonic soliton solutions on a non-BPS D$p$-brane which can be identified with BPS 
D($p-1$)-branes. In more detail, we find a kink solution whose tension is 94.9\% of that 
of a BPS D$(p-1)$-brane and a kink-antikink solution with its tension being 98.8\% of the sum 
of the tensions of a BPS D$(p-1)$-brane and a BPS anti-D$(p-1)$-brane, in the 
level truncation scheme. 

This paper is organized as follows. In section 2, after reviewing the formalism of Berkovits' 
superstring field theory, we prepare the level-expanded string field for the calculation of 
the action. In section 3, we describe the construction of the solitonic solutions and then 
compare their tensions with the expected results. In section 4 we briefly summarize our results.
Some of the details about the calculations are collected in Appendices. 

\section{Superstring Field Theory}
In this section we describe only the relevant parts of the structure of the theory for 
later calculations. For more details, see~\cite{BSZ} and chapter 3 of~\cite{KO}. 

\subsection{The action for the string field}
The Wess-Zumino-Witten--like action for the Neveu-Schwarz sector string field $\widehat{\Phi}$ 
on a non-BPS D$p$-brane of type II superstring theory is written as 
\begin{eqnarray}
\hspace{-2cm}
S&=&\frac{1}{4g_o^2}\bllk\left(e^{-\widehat{\Phi}}\widehat{Q}_B
e^{\widehat{\Phi}}\right)\left(e^{-\widehat{\Phi}}\widehat{\eta}_0
e^{\widehat{\Phi}}\right) \nonumber \\
& &{}-\int_0^1dt\left(e^{-t\widehat{\Phi}}\partial_te^{t\widehat{\Phi}}\right)
\left\{\left(e^{-t\widehat{\Phi}}\widehat{Q}_Be^{t\widehat{\Phi}}\right),
\left(e^{-t\widehat{\Phi}}\widehat{\eta}_0e^{t\widehat{\Phi}}\right)\right\}
\brrk \label{eq:A} \\ &=&\frac{1}{2g_o^2}\sum_{M,N=0}^{\infty}\frac{(-1)^N}{(M+N+2)!}
\left({M+N \atop N}\right)\bllk\left(\widehat{Q}_B\widehat{\Phi}\right)
\widehat{\Phi}^M\left(\widehat{\eta}_0\widehat{\Phi}\right)\widehat{\Phi}^N
\brrk, \nonumber
\end{eqnarray}
where 
\begin{eqnarray}
\widehat{Q}_B&=&(Q_0+Q_1+Q_2)\otimes \sigma_3, \quad \widehat{\eta}_0=\oint\frac{dz}{2\pi i}
\eta (z)\otimes\sigma_3, \label{eq:B} \\
Q_0&=&\oint\frac{dz}{2\pi i}(cT^{\mathrm{m}}+c\ \partial\xi\ \eta+cT^{\phi}+c\ \partial c\ b)(z),
\nonumber \\ Q_1&=&\oint\frac{dz}{2\pi i}\eta e^{\phi}G^{\mathrm{m}}(z), \label{eq:C} \\
Q_2&=&-\oint\frac{dz}{2\pi i}\eta\ \partial\eta\ e^{2\phi}b(z). \nonumber 
\end{eqnarray}
The BRST charge $Q_B$ was decomposed into three parts according to the $\phi$-charge of 
each term. $\sigma_3$ is the \textit{internal Chan-Paton} matrix explained below. And 
$\xi,\eta,\phi$ are the bosonized superconformal ghosts defined as 
\begin{equation}
\beta=e^{-\phi}\partial\xi,\quad \gamma=\eta e^{\phi}. \label{eq:D}
\end{equation}
Notice the \textit{orderings} of the factors above because $\xi,\eta,e^{\phi}$ are all fermionic. 
Some basic properties of world-sheet fields are summarized in Table~\ref{tab:A}. 
\begin{table}[htbp]
\begin{center}
	\begin{tabular}{|l||c|c||c|c||c|c||c|c|c||c|c|}
	\hline
	holomorphic field & $\partial X^{\mu}$ & $\psi^{\mu}$ & $b$ & $c$ 
	& $\beta$ & $\gamma$ & $e^{\ell\phi}$ & $\xi$ & $\eta$ & $T$ & $G$ \\
	\hline \hline
	conformal weight $h$ &1&1/2&2&$-1$&3/2&$-1/2$&$-\frac{1}{2}\ell^2-\ell$
	&0&1&2&3/2 \\
	\hline
	ghost number $\gh$ &0&0& $-1$ & $+1$ & $-1$ & $+1$ &0& $-1$ & $+1$ &0&0 \\
	\hline
	picture number $\pic$ &0&0&0&0&0&0& $\ell$ & $+1$ & $-1$ &0&0 \\
	\hline
	world-sheet statistics &B&F&F&F&B&B& ${\mathrm{F} (\ell:\mathrm{odd}) \atop 
	\mathrm{B} (\ell:\mathrm{even})}$  &F&F&B&F \\
	\hline
	\end{tabular}
	\caption{Some properties of the fields on an $\cN=1$ superstring 
	 world-sheet.}
	\label{tab:A}
\end{center}
\end{table}

\smallskip
The operator product expansions (OPEs) among various fields on the disk boundary are 
\begin{eqnarray}
e^{ip\cdot X(z)}e^{iq\cdot X(w)}&\sim& |z-w|^{2\ap p\cdot q}:e^{ip\cdot X(z)}e^{iq\cdot X(w)}:,
\label{eq:E1} \\ \partial X^{\mu}(z)e^{ip\cdot X(w)}&\sim& -2i\ap \frac{p^{\mu}}{z-w}e^{ip\cdot X(w)}, 
\label{eq:E2} \\ \partial X^{\mu}(z)\partial X^{\nu}(w)&\sim&-2\ap\frac{\eta^{\mu\nu}}{(z-w)^2}, 
\label{eq:E3} \\ \psi^{\mu}(z)\psi^{\nu}(w)&\sim& \frac{\eta^{\mu\nu}}{z-w}, \label{eq:E4} \\
b(z)c(w)&\sim&\frac{1}{z-w},\quad \xi(z)\eta(w)\sim\frac{1}{z-w}, \label{eq:E5} \\
e^{\ell\phi(z)}e^{m\phi(w)}&\sim& (z-w)^{-\ell m}:e^{\ell\phi(z)}e^{m\phi(w)}:, \label{eq:E6} \\
\partial\phi(z)e^{\ell\phi(w)}&\sim&-\frac{\ell}{z-w}e^{\ell\phi(w)}. \label{eq:E7}
\end{eqnarray}
The energy momentum tensors and the supercurrent are 
\begin{eqnarray}
T^{\mathrm{m}}(z)&=&-\frac{1}{4\ap}:\partial X^{\mu}\partial X_{\mu}(z):-\frac{1}{2}:\psi^{\mu}
\partial\psi_{\mu}(z):, \nonumber \\
G^{\mathrm{m}}(z)&=&\frac{i}{\sqrt{2\ap}}\psi^{\mu}\partial X_{\mu}(z), \label{eq:F} \\
T^{\phi}(z)&=&-\frac{1}{2}:\partial\phi\partial\phi(z):-\partial^2\phi(z) \nonumber 
\end{eqnarray}
so that 
\begin{eqnarray}
T^{\mathrm{m}}(z)T^{\mathrm{m}}(w)&\sim&\frac{15}{2(z-w)^4}+\frac{2}{(z-w)^2}T^{\mathrm{m}}(w)
+\frac{1}{z-w}\partial T^{\mathrm{m}}(w), \label{eq:G1} \\
G^{\mathrm{m}}(z)G^{\mathrm{m}}(w)&\sim&\frac{10}{(z-w)^3}+\frac{2}{z-w}T^{\mathrm{m}}(w).
\label{eq:G2}
\end{eqnarray}

The string field is equipped with the following Chan-Paton structure 
\begin{equation}
\widehat{\Phi}=\Phi_+\otimes I+\Phi_-\otimes\sigma_1=\left(
	\begin{array}{cc}
	\Phi_+ & \Phi_- \\
	\Phi_- & \Phi_+
	\end{array}
\right) \label{eq:H}
\end{equation}
according to the $e^{\pi i F}$ eigenvalue, where $F$ measures the world-sheet fermion number. 
Due to this, the simple derivation properties 
\begin{equation}
\widehat{Q}_B\left(\widehat{\Phi}_1\widehat{\Phi}_2\right)=\left(\widehat{Q}_B\widehat{\Phi}_1
\right)\widehat{\Phi}_2+\widehat{\Phi}_1\left(\widehat{Q}_B\widehat{\Phi}_2\right),\quad 
\widehat{\eta}_0\left(\widehat{\Phi}_1\widehat{\Phi}_2\right)=\left(\widehat{\eta}_0\widehat{\Phi}_1
\right)\widehat{\Phi}_2+\widehat{\Phi}_1\left(\widehat{\eta}_0\widehat{\Phi}_2\right), \label{eq:I}
\end{equation}
of $\widehat{Q}_B$ and $\widehat{\eta}_0$ are preserved. In addition, they satisfy 
\begin{equation}
\left\{\widehat{Q}_B,\widehat{\eta}_0\right\}=0,\quad \widehat{Q}_B^2=\widehat{\eta}_0^2=0.
\label{eq:J}
\end{equation}

\medskip

Finally, the `bracket' $\llk\ldots\rrk$ in eq.(\ref{eq:A}) is defined by 
\begin{equation}
\llk\widehat{A}_1\ldots\widehat{A}_n\rrk=\mathrm{Tr}_{\mbox{\tiny{Chan-Paton}}}\left\langle
g^{(n)}_1\circ\widehat{A}_1(0)\ldots g^{(n)}_n\circ\widehat{A}_n(0)\right\rangle, \label{eq:K}
\end{equation}
where $\widehat{A}_i$'s are arbitrary vertex operators with Chan-Paton matrices. $g^{(n)}_k(z)$ 
is geometrically the conformal transformation which maps an upper half disk representing the 
$k$-th open string world-sheet to the $k$-th $\left(\frac{360}{n}\right)^{\circ}$ wedge 
belonging to the global unit disk realizing the Witten's associative gluing interaction among 
open strings, and is concretely defined to be 
\begin{equation}
g^{(n)}_k(z)=e^{\frac{2\pi i}{n}(k-1)}\left(\frac{1+iz}{1-iz}\right)^{\frac{2}{n}}\quad 
\mathrm{for} \ \ 1\le k\le n. \label{eq:L}
\end{equation}
For later convenience, we record 
\begin{eqnarray}
g^{(n)\prime}_k(0)&=&\frac{4i}{n}e^{2\pi i\frac{k-1}{n}}, \label{eq:M1} \\
g^{(n)\prime\prime}_k(0)&=&\left(\frac{4i}{n}\right)^2e^{2\pi i\frac{k-1}{n}}. \label{eq:M2}
\end{eqnarray}
$g^{(n)}_k\circ A$ appearing in eq.(\ref{eq:K}) denotes the conformal transform of $A$ 
by $g^{(n)}_k$ and is \textit{na\"{\i}vely} written as 
\begin{equation}
g^{(n)}_k\circ A(0)\simeq \left(g^{(n)\prime}_k(0)\right)^h A\Bigl(g(0)\Bigr), \label{eq:N}
\end{equation}
where we assumed that $A$ is a primary field of conformal weight $h$. A problem arises 
when $h$ is non-integral. In fact, we will later deal with vertex operators having non-zero 
momenta, typically written as $A=A_0e^{ipX}$. Dividing the weight $h$ into two parts as 
\[ h=\ap p^2+h_N, \]
where $h_N$ denotes the contribution from $A_0$, we unambiguously define the conformal transform as 
\begin{equation}
g^{(n)}_k\circ(A_0e^{ipX})(0)=\Bigg|\left(\frac{4}{n}\right)^{\ap p^2+h_N}\Bigg|e^{2\pi ih_N
\left(\frac{k-1}{n}+\frac{1}{4}\right)}A_0e^{ipX}\Bigl(g^{(n)}_k(0)\Bigr), \label{eq:O}
\end{equation}
where use was made of eq.(\ref{eq:M1}). $\langle\ldots\rangle$ in eq.(\ref{eq:K}) then represents 
the correlation function in the combined matter-ghost superconformal field theory on the unit 
disk with $n$ conformally transformed vertex operators inserted on the boundary of the disk. 
It is normalized as 
\begin{equation}
\left\langle\xi(z_1)\frac{1}{2}c\partial c\partial^2 c(z_2)e^{-2\phi(z_3)}\right\rangle =1,
\label{eq:P}
\end{equation}
which is independent of $(z_1,z_2,z_3)$ because the above insertions play the r\^ole of filling 
the zero-modes. Note that the string field theory under consideration is formulated in the 
``large" Hilbert space including the $\xi$ zero mode, and that this fact is taken into account 
in the normalization~(\ref{eq:P}). 
\smallskip

Here we write down some important properties which $\llk\ldots\rrk$ possesses\footnote{For the proofs 
of these relations see \cite{BSZ,KO}.}:
\begin{eqnarray}
\llk A_1\ldots A_{n-1}\Phi\rrk&=&\llk\Phi A_1\ldots A_{n-1}\rrk, \nonumber \\
\llk A_1\ldots A_{n-1}(Q_B\Phi)\rrk&=&-\llk(Q_B\Phi)A_1\ldots A_{n-1}\rrk , \qquad (\mathrm{cyclicity}) 
\label{eq:Q} \\ \llk A_1\ldots A_{n-1}(\eta_0\Phi)\rrk&=&-\llk(\eta_0\Phi)A_1\ldots A_{n-1}\rrk , 
\nonumber \\ \llk Q_B(\ldots)\rrk&=&\llk\eta_0(\ldots)\rrk=0. \label{eq:R}
\end{eqnarray}
In the last line of eq.(\ref{eq:A}), we have already exploited the cyclicity properties implicitly. 
Thanks to the relations~(\ref{eq:R}), we can carry out `partial integration' which technically 
facilitates the actual computations of correlators. Now, we are ready to compute the string field 
theory action~(\ref{eq:A}) if we are given an explicit representation of the string field 
$\widehat{\Phi}$ in terms of vertex operators. 

\subsection{Level expansion of the string field on a non-BPS D-brane}\label{sec:expand}
In this paper, we will search for codimension 1 soliton solutions to the equations of motion derived 
from the action~(\ref{eq:A}) in the level truncation scheme. Let us consider the open superstring 
field theory on a non-BPS D$p$-brane, and pick out one particular direction, say $X^9\equiv X$, 
tangential to the original D$p$-brane along which we will develop non-trivial field configurations. 
We denote by $\cM$ the remaining 9-dimensional manifold. Since we require the string field configuration 
to have zero momentum in any direction along $\cM$, we can consistently restrict 
the string field to the universal subspace of the full string Hilbert space when focusing on $\cM$: 
This subspace is spanned by linear 
combinations of states obtained by acting on the zero-momentum oscillator vacuum with the ghost 
oscillators $b_n,c_n,\beta_r,\gamma_r$ and the matter super-Virasoro generators $L_n^{\cM},G_r^{\cM}$. 
This subspace is universal because it has no dependence on the details of the boundary 
conformal field theory describing the D-brane~\cite{Univ}. Along the above discussions, we decompose 
the full matter super-Virasoro generators as 
\begin{equation}
L_n^{\mathrm{m}}=L_n^X+L_n^{\cM},\quad G_r^{\mathrm{m}}=G_r^X+G_r^{\cM}, \label{eq:T}
\end{equation}
the meaning of which would be obvious. But we must pay attention to the following OPE 
\begin{equation}
G^{\cM}(z)G^{\cM}(w)\sim\frac{\mathit{9}}{(z-w)^3}+\frac{2}{z-w}T^{\cM}(w). \label{eq:U}
\end{equation}
Regarding the $X$-direction, however, the string field is forced to lie outside the universal 
class, so we choose the oscillator basis instead of the universal basis in the matter sector too. 
\medskip 

Here we define the \textit{level} of a state $|\Phi,p\rangle$ with momentum $p$ to be 
\begin{equation}
\mathrm{level}=h+\frac{1}{2}=\ap p^2+h_N+\frac{1}{2}, \label{eq:S}
\end{equation}
where the conformal weight $h$ is the $L_0^{\mathrm{tot}}$ eigenvalue. Namely, we adopt the 
`modified' definition of the level introduced in~\cite{MSZ}. And the level of a component field 
is defined to be equal to that of the associated state. According to~\cite{MSZ}, we compactify 
the $X$-direction on a circle of radius $R$, so that $X\sim X+2\pi R$ and that the momentum is 
quantized as $p=n/R$ with $n$ taking (half-)integer values. 
To this end we calculate the action using the string field 
truncated up to level $\frac{3}{2}+\frac{\ap}{R^2}$. Making use of the twist 
symmetry, it turns out that component fields of oscillator level $\frac{1}{2}$ and 1 need not acquire 
non-vanishing expectation values in constructing soliton solutions~\cite{BSZ}, where by `oscillator 
level' we mean the value of the level excluding the contribution from momentum, \textit{i.e.} 
$h_N+\frac{1}{2}$. So we set these fields to zero altogether from the beginning. We further diminish the 
number of states by imposing the Feynman-Siegel--$\xi$ gauge conditions on the string field, 
\begin{equation}
b_0|\Phi\rangle=\xi_0|\Phi\rangle=0. \label{eq:V}
\end{equation}
Though these gauge conditions are valid at the linearized level as shown in~\cite{SZ} and in 
section 2.4 and 3.6 of~\cite{KO}, it is not clear whether they are also good even non-perturbatively. 
As to the lump solution\footnote{For the spatially homogeneous `closed string vacuum' solution 
of~\cite{SZ,potential}, the validity of the Feynman-Siegel gauge condition was established in~\cite{HS}.} 
in bosonic cubic string field theory found in~\cite{MSZ}, the validity of the Feynman-Siegel gauge condition 
was verified in~\cite{Mukho} by checking that the lump solution obtained \textit{in} the Feynman-Siegel 
gauge actually solves the full set of equations of motion derived from the gauge invariant (\textit{i.e.} 
not gauge-fixed) string field theory action. Here we use the gauge conditions~(\ref{eq:V}) by simply 
assuming the validity of this gauge in constructing the classical 
solutions. Legitimacy of $b_0$ gauge condition $b_0|\Phi\rangle=0$ can probably be verified in 
the same way as in~\cite{Mukho} by including the states which do not satisfy $b_0|\Phi\rangle=0$ 
(\textit{i.e.} the states containing $c_0$ oscillator) and examining whether our solutions solve 
the full set of equations of motion derived from the gauge invariant action. Getting rid of the $\xi_0$ 
gauge condition, however, is more difficult because it breaks the one-to-one correspondence 
between the states in the ``small" Hilbert space and those in the ``large" Hilbert space. 
Answering this question in this superstring field theory remains open. 
\medskip

Putting the above accounts together, we can write down the level-truncated string field in the 
following way. First, required states in the ``small" Hilbert space are  
\begin{eqnarray}
\cT_n&=&|\tilde{\Omega},n\rangle=c_1e^{-\phi(0)}e^{i\frac{n}{R}X(0)}|0\rangle \qquad \mbox{at 
oscillator level 0,} \label{eq:W1} \\ 
\cA_n&=&c_{-1}\beta_{-1/2}|\tilde{\Omega},n\rangle, \quad \cE_n=b_{-1}\gamma_{-1/2}|\tilde{\Omega},n\rangle,
\nonumber \\ \cF_n&=&G^{\cM}_{-3/2}|\tilde{\Omega},n\rangle, \quad \cG_n=\beta_{-1/2}\gamma_{-1/2}
\psi^X_{-1/2}|\tilde{\Omega},n\rangle, \label{eq:W2} \\ \cH_n&=&\psi^X_{-3/2}|\tilde{\Omega},n\rangle, 
\quad \cJ_n=\psi^X_{-1/2}\alpha_{-1}^X|\tilde{\Omega},n\rangle, \nonumber 
\end{eqnarray}
where $|0\rangle$ is the $SL_2$-invariant vacuum, and all these states have ghost number 1 and 
picture number $-1$. The last six states are at oscillator level 3/2. As mentioned earlier, 
we adopt the complete oscillator basis in the $X$-direction, whereas in the $\cM$-direction the 
universal Virasoro basis is employed. When they are translated into the vertex operator 
representation in the ``large" Hilbert space, the $\xi_0$ gauge condition~(\ref{eq:V}) allows us to 
map any state in the ``small" Hilbert space to a certain vertex operator in the ``large" one in 
a one-to-one way. That is, the rule of translation is simply given by 
\begin{eqnarray}
\cT_n=c_1e^{-\phi(0)}e^{i\frac{n}{R}X(0)}|0\rangle&\longrightarrow& ce^{-\phi}e^{i\frac{n}{R}X}
\longrightarrow T_n=\xi ce^{-\phi}e^{i\frac{n}{R}X} \label{eq:X} \\
\mathrm{state} \qquad & & \mbox{``small"} \qquad\qquad \mbox{``large"} \nonumber 
\end{eqnarray}
for the tachyon state, for example. Using this rule, the string field $\widehat{\Phi}$ takes the 
following form, 
\begin{eqnarray}
\widehat{\Phi}&=&T\otimes\sigma_1+(E+V+A)\otimes I; \label{eq:Y} \\
& &V=G+F+H+J, \nonumber 
\end{eqnarray}
where
\begin{eqnarray}
T&=&\sum_nt_nT_n=\sum_nt_n\ \xi ce^{-\phi}e^{i\frac{n}{R}X}, \nonumber \\
E&=&\sum_ne_nE_n=\sum_ne_n\ \xi\eta e^{i\frac{n}{R}X}, \nonumber \\
A&=&\sum_na_nA_n=\sum_na_n\ \xi\partial\xi c\partial^2ce^{-2\phi}e^{i\frac{n}{R}X}, \nonumber \\
G&=&\sum_ng_nG_n=\sum_ng_n\ \xi c(\partial e^{-\phi})\psi^Xe^{i\frac{n}{R}X}, \label{eq:Z} \\
F&=&\sum_nf_nF_n=\sum_nf_n\ \xi ce^{-\phi}G^{\cM}e^{i\frac{n}{R}X}, \nonumber \\
H&=&\sum_nh_nH_n=\sum_nh_n\ \xi ce^{-\phi}(\partial \psi^X)e^{i\frac{n}{R}X}, \nonumber \\
J&=&\sum_nj_nJ_n=\frac{i}{\sqrt{\ap}}\sum_nj_n\ \xi ce^{-\phi}\psi^X\partial Xe^{i\frac{n}{R}X}. 
\nonumber 
\end{eqnarray}
Since $\xi$ has ghost number $-1$ and picture number 1, the string field $\widehat{\Phi}$ is 
expressed by vertex operators of ghost number 0 and picture number 0, as required. The tachyon 
vertex operator $T$ is tensored by the intenal Chan-Paton matrix $\sigma_1$ because the Neveu-Schwarz 
sector ground state $|\tilde{\Omega},n\rangle$ itself is in the GSO($-$) sector, while the other 
states appearing in~(\ref{eq:W2}) lie in the GSO(+) sector so that they are multiplied by $I$. 
Since $E,A,G,H$ and $J$ are not conformal primary fields, their conformal properties are 
complicated: we record them, as well as the action of the BRST charge on the vertex operators, 
in Appendix~\ref{sec:appA}. 

\subsection{Calculation of the action}
Now that we have explained all ingredients, we can in principle calculate the string field 
theory action~(\ref{eq:A}) and express it in terms of the component fields 
$\{t_n,e_n,a_n,g_n,f_n,h_n,j_n\}$ by substituting the string field~(\ref{eq:Y}) into the 
action~(\ref{eq:A}), carrying out the conformal transformations, acting on the vertex 
operators with $Q_B$ and 
$\eta_0$, and evaluating the correlators on the disk using the OPEs~(\ref{eq:E1})--(\ref{eq:E7}), 
(\ref{eq:U}) and the normalization~(\ref{eq:P}). But unfortunately, computations are tedious and 
even the final expressions of the action are awfully lengthy. So we present only some representative 
terms in Appendix~\ref{sec:appB}. 

\section{Solutions and Their Properties}
In this section, we search for solutions which non-trivially depend on the coordinate in the $X$-direction 
by extremizing the action written in terms of a finite number of component fields. For a technical reason, 
we deal with the case of a single kink and the case of a kink-antikink pair separately 
in subsection 3.1 and 3.2, respectively. 

\subsection{Kink solution}
First of all, we must define `what is the kink solution'. Since the matter part of the 
tachyon vertex operator has the form 
\[T^{\mathrm{matter}}=\sum_nt_ne^{i\frac{n}{R}X}, \]
we can naturally consider the following function 
\begin{equation}
t(x)=\sum_nt_ne^{i\frac{n}{R}x} \label{eq:BA}
\end{equation}
of the center-of-mass coordinate $x$ as representing the profile of the tachyon field on the 
original non-BPS D$p$-brane, as is done in~\cite{MSZ}. The reality condition $t(x)^*=t(x)$ of the 
tachyon field demands that its Fourier components $\{t_n\}$ should satisfy 
\begin{equation}
t^*_n=t_{-n}. \label{eq:BB}
\end{equation}
Since the tachyon potential on a non-BPS D-brane has doubly-degenerate minima $\pm\overline{t}$, 
we could obviously define a tachyonic kink configuration, if the $X$-direction was non-compact 
$\aaru$, to be 
\begin{equation}
t(x)\longrightarrow\left\{
	\begin{array}{ccc}
	\overline{t} & \mathrm{for} & x\to +\infty \\
	-\overline{t} & \mathrm{for} & x\to -\infty
	\end{array} 
\right. , \label{eq:BC}
\end{equation}
and similarly an antikink configuration with the signs reversed. But now, we are compactifying 
the $X$-direction on a circle of radius $R$, so that $x\in \cI =[-\pi R,\pi R]$. In this case, 
we refer to a configuration which obeys 
\begin{equation}
t(x)\sim\left\{
	\begin{array}{ccc}
	\overline{t} & \mathrm{near} & x=\pi R \\
	-\overline{t} & \mathrm{near} & x=-\pi R
	\end{array}
\right. \label{eq:BD}
\end{equation}
and is a monotonically increasing function of $x\in\cI$ as a tachyonic \textit{kink}. 
From eq.(\ref{eq:BD}), we are immediately aware that the tachyon field $t(x)$ must satisfy 
\textit{antiperiodic} boundary condition. This is achieved by letting the discrete momentum $n$ 
take value in $\zetto +\frac{1}{2}$. Physically, it corresponds to turning on a $\zetto_2$ Wilson 
line along the circle so that the boundary condition is twisted~\cite{SO32,MSZ}. For ease of 
calculations, we will look for a kink solution whose profile is an odd function of $x\in\cI$. This 
condition gives 
\begin{equation}
t(x)=-t(-x) \quad \longrightarrow \quad t_{-n}=-t_n. \label{eq:BE}
\end{equation}
This constraint, combined with~(\ref{eq:BB}), means that every $t_n$ must be purely imaginary, 
$t_n^*=-t_n$. Hence we set 
\begin{equation}
t_n=-t_{-n}=\frac{1}{2i}\tau_n \label{eq:BF}
\end{equation}
with $\tau_n$ real so that 
\begin{equation}
t(x)=\sum_{n\in\zetto+\frac{1}{2}}t_ne^{i\frac{n}{R}x}=\sum_{n\in\zetto_{\scriptscriptstyle{\ge 0}}
+\frac{1}{2}}\tau_n \sin\frac{n}{R}x. \label{eq:BG}
\end{equation}

\subsubsection{Kink solution in purely tachyonic superstring field theory}
\begin{figure}[htbp]
	\begin{center}
	\includegraphics{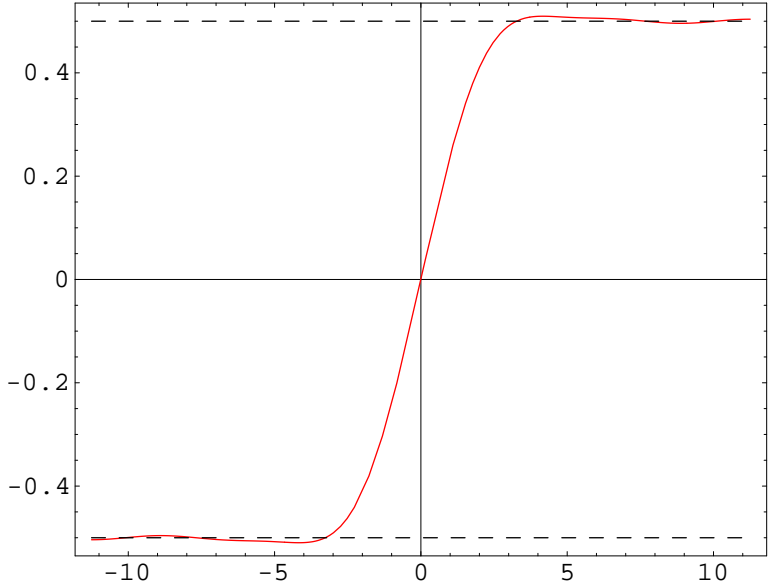}
	\end{center}
	\caption{The solid line shows a plot of $t(x)$ found at the purely tachyonic level. 
	The dashed lines indicate the vacuum expectation values $\pm\overline{t}=\pm 0.5$ of 
	the tachyon field at the minima of the level $(0,0)$ tachyon potential.}
	\label{fig:ptkink}
\end{figure}

By `purely tachyonic superstring field theory', we mean the one in which the string field 
$\widehat{\Phi}$ contains only the tachyon state and its harmonics, namely the superstring 
field theory truncated to the oscillator level 0. In this theory, the action involves only 
those terms which are quadratic and quartic in the tachyon field, so that the action can 
be calculated fairly easily. The result is 
\begin{eqnarray}
2g_o^2S_T&=&\frac{1}{2}\bllk\left(\widehat{Q}_B\widehat{T}\right)\left(\widehat{\eta}_0
\widehat{T}\right)\brrk+\frac{1}{12}\left(\bllk\left(\widehat{Q}_B\widehat{T}\right)\widehat{T}^2
\left(\widehat{\eta}_0\widehat{T}\right)\brrk-\bllk\left(\widehat{Q}_B\widehat{T}\right)\widehat{T}
\left(\widehat{\eta}_0\widehat{T}\right)\widehat{T}\brrk\right) \nonumber \\
&=&2\pi R V_p\Biggl\{\sum\left(-\frac{\ap}{R^2}n^2+\frac{1}{2}\right)t_nt_{-n} \label{eq:BH} \\
& & -\sum\delta_{n_1+n_2+n_3+n_4}t_{n_1}t_{n_2}t_{n_3}t_{n_4}\exp\left[\frac{\ap}{R^2}\log 2
\left((n_1+n_3)^2-\sum_{I=1}^4(n_I)^2\right)\right]\Biggr\}, \nonumber 
\end{eqnarray}
where $V_p$ is the volume of $p$-dimensional subspace of the original D$p$-brane perpendicular 
to the $X$-direction. The details of the calculations are shown in Appendix~\ref{sec:appB}. 
In order to find a numerical solution, we must assign a numerical value to 
the radius $R$. In this subsection, we choose 
\begin{equation}
R=\sqrt{\frac{77}{6}\ap} \label{eq:BI}
\end{equation}
for a reason mentioned in the next sub-subsection. Given this value, the level of each Fourier 
mode $t_n$ of the tachyon field is found to be 
\begin{equation}
\mathrm{level}(t_n)=\mathrm{level}(\tau_n)=\frac{\ap}{R^2}n^2=\frac{6}{77}n^2, \label{eq:BJ}
\end{equation}
and we keep only those modes with level equal to or less than $\frac{243}{154}=\frac{3}{2}+
\frac{6}{77}$. In other words, the summation is taken over $|n|\le 9/2$. Furthermore, we 
truncate the action at level $243/77$, which means that in the quartic interaction terms in 
eq.(\ref{eq:BH}) we discard those terms with level 
\[ \frac{6}{77}(n_1^2+n_2^2+n_3^2+n_4^2) \]
exceeding $243/77$. That is, we adopt the level $(\frac{3}{2}+\frac{\ap}{R^2},3+2\frac{\ap}{R^2})=
(\frac{243}{154},\frac{243}{77})$ approximation to the action. This procedure gives the following 
action,\footnote{In the original version of this paper, we had dropped the factor of $-\frac{1}{\pi^2}$ 
in eq.(\ref{eq:BK}).}
\begin{eqnarray}
2g_o^2S_T^{\left(\frac{243}{154},\frac{243}{77}\right)}&=&-\frac{2\pi RV_p}{\pi^2}
(-2.37 \tau_{1/2}^2+3.57 \tau_{1/2}^4
-4.43 \tau_{1/2}^3\tau_{3/2}-1.60 \tau_{3/2}^2+12.42 \tau_{1/2}^2\tau_{3/2}^2 \nonumber \\ 
& &+2.75 \tau_{3/2}^4 -10.74 \tau_{1/2}^2\tau_{3/2}\tau_{5/2}+10.18\tau_{1/2}\tau_{3/2}^2
\tau_{5/2}-0.0641 \tau_{5/2}^2 
\nonumber \\ & &+9.46\tau_{1/2}^2 \tau_{5/2}^2+8.73\tau_{3/2}^2\tau_{5/2}^2+1.87 \tau_{5/2}^4-8.09 
\tau_{1/2}\tau_{3/2}^2\tau_{7/2}-7.62\tau_{1/2}^2\tau_{5/2}\tau_{7/2}\nonumber \\ & &+14.83
\tau_{1/2}\tau_{3/2}\tau_{5/2}\tau_{7/2}+6.85\tau_{3/2}\tau_{5/2}^2\tau_{7/2}+2.24\tau_{7/2}^2
+6.39 \tau_{1/2}^2\tau_{7/2}^2 \nonumber \\ & &+6.29\tau_{3/2}^2\tau_{7/2}^2+6.05\tau_{5/2}^2\tau_{7/2}^2 
-1.87\tau_{3/2}^3\tau_{9/2}-10.80\tau_{1/2}\tau_{3/2}\tau_{5/2}\tau_{9/2} \nonumber \\ & & 
+5.26\tau_{1/2}\tau_{5/2}^2\tau_{9/2} -4.82\tau_{1/2}^2\tau_{7/2}\tau_{9/2}
+9.75\tau_{1/2}\tau_{3/2}\tau_{7/2}\tau_{9/2}
+5.32\tau_{9/2}^2) \label{eq:BK}
\end{eqnarray}
where we used eq.(\ref{eq:BF}). Since the action has been rewritten as a simple quartic polynomial 
of five variables $(\tau_{1/2},\tau_{3/2},\tau_{5/2},\tau_{7/2},\tau_{9/2})$, we can easily 
extremize it by solving simultaneous equations of motion $\{\partial S_T/\partial\tau_n=0\}$ 
numerically. With a suitable choice of initial values for $\tau_n$'s, our numerical algorithm 
has converged to 
\begin{eqnarray}
\Phi_k&=&\Bigl\{\tau_{1/2}=0.626558, \tau_{3/2}=0.180632, \tau_{5/2}=0.0745727, \nonumber \\
& &\tau_{7/2}=0.0264652,
\tau_{9/2}=0.00983679\Bigr\}. \label{eq:BL}
\end{eqnarray}
Substituting these values into eq.(\ref{eq:BG}) we can find the tachyon profile, which is 
plotted in Figure~\ref{fig:ptkink}. 
This figure nicely shows that the solution~(\ref{eq:BL}) corresponds to the desired tachyonic 
kink configuration. In particular, the tachyon field correctly approaches the vacuum values 
in the asymptotic region $x\sim\pm\pi R$.
\medskip

The next step is to calculate the tension of the kink solution. First, we reexpress the open string 
coupling constant $g_o$ defined for open superstrings on the original non-BPS D$p$-brane in terms of 
the tension $\tilde{\tau}_p$ of the non-BPS D$p$-brane. This relation has already been worked out 
in~\cite{BSZ} as 
\begin{equation}
\tilde{\tau}_p=\frac{1}{2\pi^2g_o^2}. \label{eq:BM}
\end{equation}
Using it, the purely tachyonic action~(\ref{eq:BH}) is rewritten as 
\begin{equation}
S_T=2\pi RV_p\tilde{\tau}_p\pi^2\left\{\sum\left(-\frac{\ap}{R^2}n^2+\frac{1}{2}\right)t_nt_{-n}
+\ldots\right\}. \label{eq:BN}
\end{equation}
Since the action on a non-BPS D$p$-brane is proportional to the $(p+1)$-dimensional volume 
$2\pi RV_p$ and the brane tension $\tilde{\tau}_p$, we generally define 
\begin{equation}
f(\Phi)=-\frac{1}{(2\pi RV_p)\tilde{\tau}_p}S(\Phi)=-\frac{\pi^2}{2\pi RV_p}2g_o^2S(\Phi). 
\label{eq:BO}
\end{equation}
As the additive normalization of the string field theory action~(\ref{eq:A}) is given by $S(0)=0$, 
the Sen's conjecture about the brane annihilation is stated as $f(\Phi_0)=-1$, where $\Phi_0$ 
denotes the `closed string vacuum' configuration. In order to calculate the kink tension, however, 
we have to add a suitable constant term to the tachyon potential such that the energy density at 
the bottom of the potential should vanish. At the purely tachyonic level, this can be done by shifting 
$f(\Phi)$ by $f^{(0,0)}(\Phi_0)\simeq -0.617$, where $f^{(0,0)}$ is the level $(0,0)$ tachyon 
potential~\cite{NS}. Then the value of the shifted action evaluated 
on the kink solution~(\ref{eq:BL}) becomes 
\begin{equation}
\cS_T^{\left(\frac{243}{154},\frac{243}{77}\right)}=2\pi RV_p\tilde{\tau}_p\left(f^{(0,0)}(\Phi_0)
-f^{\left(\frac{243}{154},\frac{243}{77}\right)}(\Phi_k)\right)\equiv -V_p\cT_{p-1}, \label{eq:BP}
\end{equation}
where we have denoted the kink tension by $\cT_{p-1}$. Since it is conjectured (and in fact verified 
in the context of boundary superstring field theory~\cite{KMM2}) that a tachyonic kink configuration 
on a non-BPS D$p$-brane is identified with a BPS D$(p-1)$-brane with tension $\tau_{p-1}=2\pi\sqrt{\ap}
\tilde{\tau}_p/\sqrt{2}$, we need to calculate the ratio 
\begin{equation}
r=\frac{\cT_{p-1}}{\tau_{p-1}}=\frac{\sqrt{2}\cT_{p-1}}{2\pi\sqrt{\ap}\tilde{\tau}_p}=\sqrt{2}
\frac{R}{\sqrt{\ap}}\left(f^{\left(\frac{243}{154},\frac{243}{77}\right)}(\Phi_k)-f^{(0,0)}
(\Phi_0)\right) \label{eq:BQ}
\end{equation}
and to see whether the value of $r$ is in the vicinity of unity. Putting the numerical 
values~(\ref{eq:BL}) into the above expression and evaluating it, we obtain 
\begin{equation}
r\simeq 0.640. \label{eq:BZ}
\end{equation}
Thus we conclude that the tension of the kink solution in the purely tachyonic superstring 
field theory is about 64\% of the expected answer, for a particular choice~(\ref{eq:BI}) of 
radius. Such a degree of accuracy, which is slightly better than that of the depth of the 
tachyon potential, is quite reasonable when we compare it with the result of the lump solutions in 
the bosonic cubic string field theory~\cite{MSZ}. Although it is interesting 
to see whether or not the value of $r$ is approximately independent of the chosen values of the 
radius $R$, we leave it for future studies.

\subsubsection{Including level 3/2 fields}
We have seen in the last sub-subsection that using the purely tachyonic theory we qualitatively 
succeed in constructing a tachyonic kink solution. Nevertheless, the value $r\simeq 0.640$ is not 
so satisfactory. In this sub-subsection we will be looking for a kink solution with a higher 
accuracy by including oscillator level 3/2 fields. 
\medskip

We begin by determining the boundary conditions, or modings, of component fields. 
By definition of the tachyonic kink, the tachyon field $t(x)$ must obey the antiperiodic boundary 
condition. Therefore, the tachyon field is again expanded in half-integral modes. Other fields 
at oscillator level 3/2, however, turn out to satisfy the \textit{periodic} boundary 
condition.\footnote{I thank T. Takayanagi for pointing this out.} To see this, let us regard the 
non-BPS D$p$-brane and various fields on it as being obtained by modding out the coincident 
D$p$-brane--anti-D$p$-brane system by the action of $(-1)^{F_L}$~\cite{4207}. Then the tachyonic mode 
arises in the spectrum of anti-GSO-projected $p$-$\bar{p}$ strings. Turning on a $\zetto_2$ Wilson 
line along the $X$-direction of one of the two branes, the boundary condition of the tachyon field, 
which is in the bifundamental representation of the gauge group $U(1)\times U(1)$, is reversed. 
On the other hand, level 3/2 fields living in the GSO(+) sector all come from 
$p$-$p$ or $\bar{p}$-$\bar{p}$ strings. Since such fields are neutral under the gauge group, their 
boundary conditions are not affected by the existence of the $\zetto_2$ Wilson line. As a result, 
these fields have integer modes. 
\smallskip

\begin{figure}[htbp]
	\begin{center}
	\includegraphics{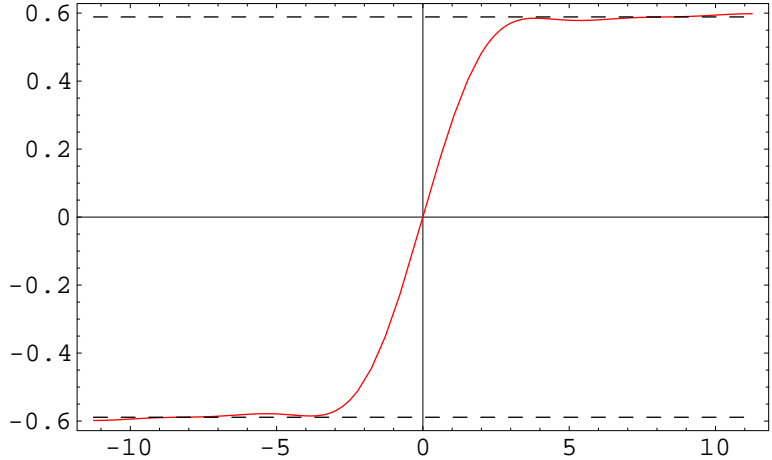}
	\end{center}
	\caption{The solid line shows a plot of $t(x)$ found in the level $\left(\frac{243}{154},
	\frac{243}{77}\right)$ truncated superstring field theory. 
	The dashed lines indicate the vacuum expectation values $\pm\overline{t}=\pm 0.58882$ of 
	the tachyon field at the minima of the level $\left(\frac{3}{2},3\right)$ tachyon potential.}
	\label{fig:sftkink}
\end{figure}

Now we proceed to level expansion of the string field. For $R=\sqrt{77\ap /6}$ the first harmonics 
($|n|=1$ modes) of the oscillator level 3/2 fields are at level 
\[ \frac{3}{2}+\frac{\ap}{R^2}=\frac{243}{154}. \]
As we have already seen before, this value coincides with the level number of the $|n|=9/2$ modes 
of the tachyon field: This is why we have chosen the strangely-looking value $R=\sqrt{77\ap /6}$ 
as the radius. Truncating the string field and the action at level $\left(\frac{243}{154},
\frac{243}{77}\right)$, one obtains the approximate action as a sixth-order polynomial 
of 23 variables, though its full expression is quite long. We have searched 
for a stationary point of this action by solving numerically 23 simultaneous equations of motion 
obtained by differentiating the action with respect to the 23 variables. Starting from 
the results~(\ref{eq:BL}) obtained in the purely tachyonic theory, we have reached the 
following set of values using our algorithm, 
\begin{equation}
\Phi_k=\left\{
	\begin{array}{lll}
	\tau_{1/2}=0.733584, & \tau_{3/2}=0.201643, & \tau_{5/2}=0.0915225, \\ 
	\tau_{7/2}=0.0363576, & 
	\tau_{9/2}=0.0111283, & \\ e_0=0.0829011, & e_1=-0.00933999, & e_{-1}=-0.00933999, \\
	a_0=0.0445132, & a_1=-0.00978734, & a_{-1}=-0.00978734, \\ f_0=0.00996780, & 
	f_1=-0.00219249, & f_{-1}=-0.00219249, \\ g_0=-2.53\times 10^{-19}, & g_1=-0.000979490, & 
	g_{-1}=0.000979490, \\ h_0=2.15\times 10^{-19}, & h_1=0.00251080, & h_{-1}=-0.00251080, \\
	j_0=-0.00416412, & j_1=-0.0103785, & j_{-1}=-0.0103785 
	\end{array}
\right\}. \label{eq:BR}
\end{equation}
The tachyon profile~(\ref{eq:BG}) is plotted in Figure~\ref{fig:sftkink}. 
Though we can hardly distinguish it from Figure~\ref{fig:ptkink}, one should note that the dashed lines 
in Figure~\ref{fig:sftkink} signify the vacuum values of the tachyon field at the minima of the 
level $\left(\frac{3}{2},3\right)$ tachyon potential. The fact that the tachyon field representing the 
kink solution correctly asymptotes to the vacuum values suggests that our action, as well as the 
superstring field theory, is right. Moreover, the tension of this kink solution can be calculated by 
substituting the values~(\ref{eq:BR}) into the formula 
\begin{equation}
r=\sqrt{2}\frac{R}{\sqrt{\ap}}\left(f^{\left(\frac{243}{154},\frac{243}{77}\right)}(\Phi_k)-
f^{\left(\frac{3}{2},3\right)}(\Phi_0)\right), \label{eq:BS}
\end{equation}
where $f(\Phi)$ is defined in eq.(\ref{eq:BO}) and the depth of the tachyon potential 
at level $\left(\frac{3}{2},3\right)$ is given by~\cite{BSZ} 
\[ f^{\left(\frac{3}{2},3\right)}(\Phi_0)\simeq -0.854. \]
The resulting value of the ratio $r$ is found to be 
\[ r\simeq 0.949. \]
We can say that this value (about 95\% of the expected one) is considerably better than the 
previous one. 

\subsubsection{An extra solution: triple brane}
\begin{figure}[htbp]
	\begin{center}
	\includegraphics{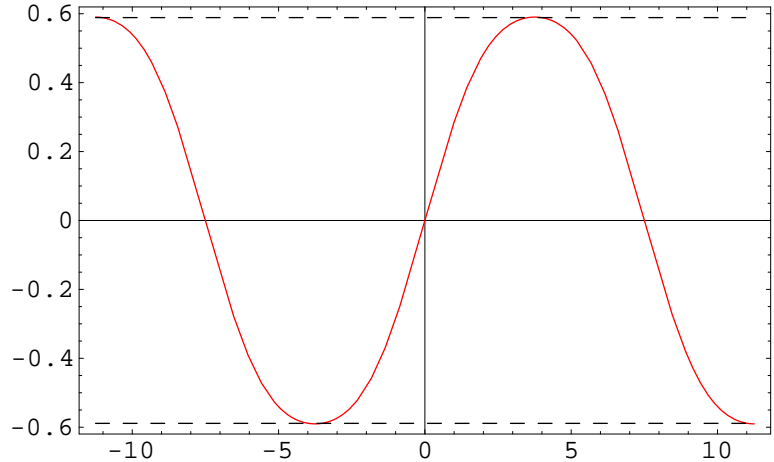}
	\end{center}
	\caption{The solid line shows a plot of $t(x)$ in the `triple brane' solution.}
	\label{fig:triple}
\end{figure}

In the course of the search for the kink solution, we have encountered a brane-antibrane-brane 
configuration (Figure~\ref{fig:triple}) whose tension is about 92\% of three times the BPS 
D$(p-1)$-brane tension. 

\subsection{Kink-antikink pair}
In this subsection we will repeat essentially the same analysis as in the last subsection, 
with $t(x)$ replaced by an even function of $x\in\cI$ satisfying the periodic boundary 
condition. $t(x)=t(-x)$ combined with the reality condition~(\ref{eq:BB}) implies 
\begin{equation}
t_n=t_{-n}=t_n^*\equiv \tau_n. \label{eq:BT}
\end{equation}
Hence the tachyon profile is now given by 
\begin{equation}
t(x)=\sum_{n\in\zetto}t_ne^{i\frac{n}{R}x}=\tau_0+2\sum_{n=1}^{(\infty)}\tau_n\cos\frac{n}{R}x. 
\label{eq:BU}
\end{equation}

\subsubsection{Lump-like solution in purely tachyonic superstring field theory}
\begin{figure}[htbp]
	\begin{center}
	\includegraphics{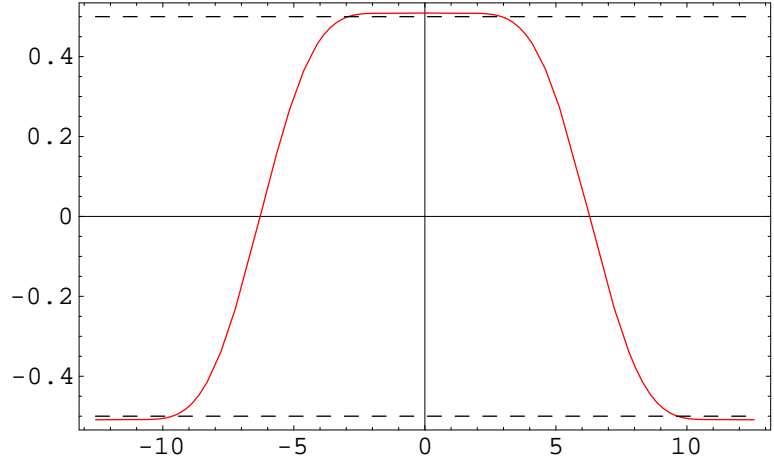}
	\end{center}
	\caption{The solid line shows a plot of $t(x)$ found at the purely tachyonic level. 
	The dashed lines indicate the vacuum expectation values $\pm\overline{t}=\pm 0.5$ of 
	the tachyon field at the minima of the level $(0,0)$ tachyon potential.}
	\label{fig:ptlump}
\end{figure}

In this subsection our choice of radius\footnote{In~\cite{MSZ} the authors avoided using rational 
values for $R/\sqrt{\ap}$ in order for the Verma module to agree with the total space spanned by 
the complete oscillator basis in the nonzero momentum sectors. This consideration does not matter 
to us here because we are adopting the oscillator basis~(\ref{eq:W2}) in the $X$-direction 
from beginning to end.} is 
\begin{equation}
R=4\sqrt{\ap} \label{eq:BV}
\end{equation}
for the same reason as mentioned in the previous subsection. Then the level of the tachyon field 
becomes 
\begin{equation}
\mathrm{level}(t_n)=\mathrm{level}(\tau_n)=\frac{\ap}{R^2}n^2=\frac{n^2}{16}. \label{eq:BW}
\end{equation}
Truncating the field and the action at level $\left(\frac{25}{16},\frac{25}{8}\right)$ and solving 
the equations of motion, we have obtained the following solution 
\begin{equation}
\Phi_{\ell}=\Bigl\{\tau_0=0. , \tau_1=0.301978,  \tau_2=0. ,  \tau_3=-0.0560825,  
\tau_4=0. ,  \tau_5=0.00860546\Bigr\} \label{eq:BX}
\end{equation}
with the profile shown in Figure~\ref{fig:ptlump}. 
Since this lump-like solution can be interpreted as 
representing a kink-antikink pair, we should compare the tension of the solution, calculated as in 
eq.(\ref{eq:BP}), with the sum of the tensions of a BPS D$(p-1)$-brane and a BPS 
anti-D$(p-1)$-brane.\footnote{One may think that the attractive force between a brane and an antibrane 
lowers the energy of the total system: This issue will be discussed in sub-subsection~\ref{sec:move}.} 
Hence the ratio $r$~(\ref{eq:BQ}) in this case should be modified as 
\begin{equation}
r=\frac{\cT_{p-1}}{2\times \tau_{p-1}}=\frac{1}{\sqrt{2}}\frac{R}{\sqrt{\ap}}\left(f^{\left(\frac{25}{16},
\frac{25}{8}\right)}(\Phi_{\ell})-f^{(0,0)}(\Phi_0)\right). \label{eq:BY}
\end{equation}
Substituting the numerical 
values~(\ref{eq:BX}) into the above formula, we get 
\[ r\simeq 0.639, \]
which happens to be very close to the previous result~(\ref{eq:BZ}) for a kink. 

\subsubsection{Including level 3/2 fields}
\begin{figure}[htbp]
	\begin{center}
	\includegraphics{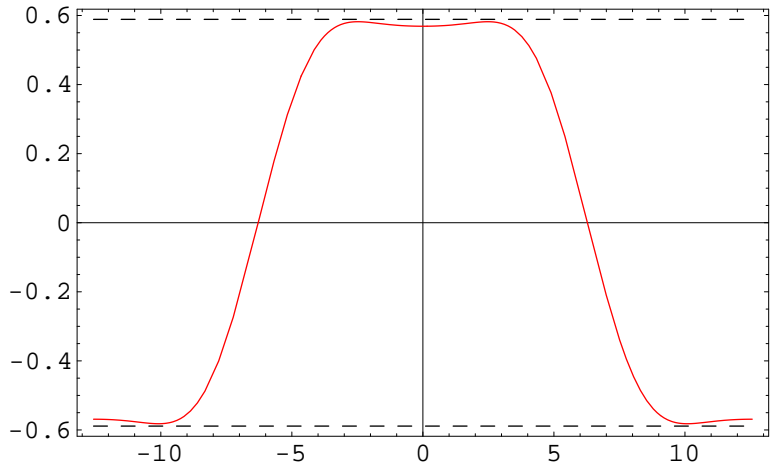}
	\end{center}
	\caption{The solid line shows a plot of $t(x)$ found in the level $\left(\frac{25}{16},
	\frac{25}{8}\right)$ truncated superstring field theory. 
	The dashed lines indicate the vacuum expectation values $\pm\overline{t}=\pm 0.58882$ of 
	the tachyon field at the minima of the level $\left(\frac{3}{2},3\right)$ tachyon potential. 
	$t(x)=0$ occurs at $x=\pm 2\pi$ for $R=4$ (in units of $\sqrt{\ap}$).}
	\label{fig:sftlump}
\end{figure}

For $R=4\sqrt{\ap}$, the level of the first harmonics ($|n|=1$) of the oscillator level 3/2 
fields is 
\[ \frac{3}{2}+\frac{\ap}{R^2}=\frac{25}{16}, \]
which coincides with that of the $|n|=5$ modes of the tachyon field. By adopting 
the level $\left(\frac{25}{16},
\frac{25}{8}\right)$ approximation to the action, it can be written as a polynomial of 24 variables. 
Solving the equations of motion gives 
\begin{equation}
\Phi_{\ell}=\left\{
	\begin{array}{lll}
	\tau_0=-9.06\times 10^{-12}, & \tau_1=0.344976, & \tau_2=6.77\times 10^{-12}, \\ 
	\tau_3=-0.0703883, & \tau_4=-2.79\times 10^{-12}, & \tau_5=0.00985764, \\ 
	e_0=0.0752118, & a_0=0.0348621, & f_0=0.00760330, \\ j_0=-0.0146609, & 
	|\mbox{other 14 modes}|<10^{-12} & 
	\end{array}
\right\}. \label{eq:CA}
\end{equation}
Note that the above solution says $\tau_0,\tau_2,\tau_4$ are practically \textit{zero}. It then 
follows that the tachyon profile~(\ref{eq:BU}) crosses the horizontal axis at $x=\pm\pi R/2$, 
as shown in Figure~\ref{fig:sftlump}. 
The physical meaning of this fact is, interestingly, that a D$(p-1)$-brane and an anti-D$(p-1)$-brane, 
constructed as a kink and an antikink on a non-BPS D$p$-brane respectively, are located at diametrically 
opposite points of the circle in the compactified $X$-direction. What we have to mention is that our 
numerical algorithm automatically converges to the solution~(\ref{eq:CA}) even if we give non-zero 
initial values to $\tau_0,\tau_2,\tau_4$. To be more convinced that this solution really corresponds 
to the brane-antibrane pair, we calculate the tension of the solution using the formula~(\ref{eq:BY}) 
with $f^{(0,0)}(\Phi_0)$ replaced by $f^{\left(\frac{3}{2},3\right)}(\Phi_0)$. Marvelously, the 
resulting value of the ratio has turned out to be 
\[ r\simeq 0.988. \]
That is to say, the tension of the solution~(\ref{eq:CA}) agrees with the expected answer to 
an accuracy of about 1.2\%!

\subsubsection{Deformation of the solution}\label{sec:move}
\begin{figure}[htbp]
	\begin{center}
	\includegraphics{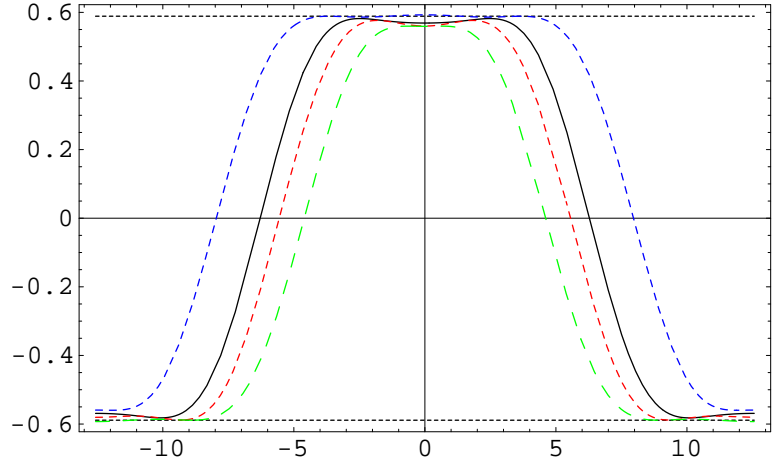}
	\end{center}
	\caption{The solid line shows a plot of $t(x)$ in the original kink-antikink solution. 
	The dashed curves represent the translated kink-antikink pairs obtained by setting 
	$\tau_2=0.1, 0.05, -0.1$ from below.}
	\label{fig:move}
\end{figure}

\begin{figure}[htbp]
	\begin{center}
	\includegraphics{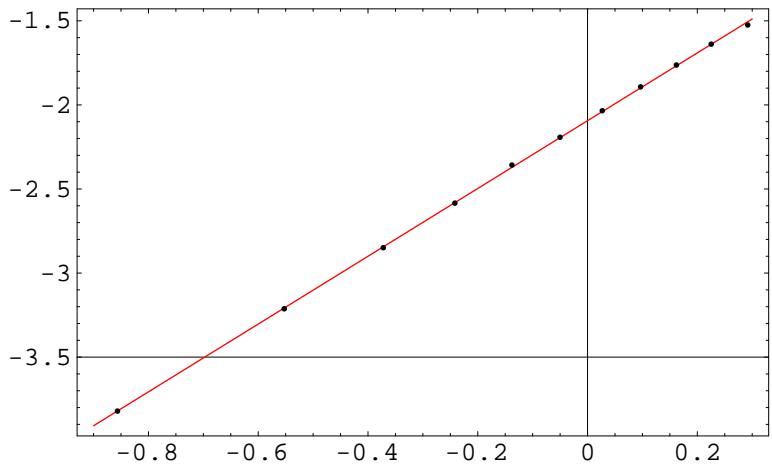}
	\end{center}
	\caption{This figure shows the relation between the ratio $r$ and the half-distance $x_0$. 
	The horizontal axis represents the value of $\log_{10}(2\pi -x_0)$, while the vertical 
	axis $\log_{10}(0.987919-r)$. The dots are taken from Table~\ref{tab:B}, and the solid line 
	shows the plot of the best fit~(\ref{eq:CB}).}
	\label{fig:fitting}
\end{figure}

Since we have constructed the multi-brane solution, we want to study 
the relative motion of constituent branes. 
(The center-of-mass motion was frozen out when we set $t(x)$ to an even 
function of $x$.) Here we insist that the interaction between a kink and an antikink should not 
be incorporated into the classical solutions of superstring field theory. The reason is as follows: 
The interaction between two D-branes comes about from the exchange of closed strings. From the 
point of view of open strings, this corresponds to a 1-loop diagram with no vertex operators inserted. 
Such loop effects of open strings, however, are not taken into account in superstring field theory 
at the \textit{classical} level because $\llk\ldots\rrk$ in the action are evaluated as \textit{disk} 
correlation functions among the vertex operators inserted. Loop corrections could be calculated if we 
expanded the string field $\widehat{\Phi}$ around our brane solution $\widehat{\Phi}_{\ell}$ as 
$\widehat{\Phi}=\widehat{\Phi}_{\ell}+\widehat{\Psi}$ and applied the \textit{quantum} superstring 
field theory to the fluctuation field $\widehat{\Psi}$. Provided that our claim is true, there will 
be a flat direction in the potential in which the branes can move freely. As can be seen from the fact, 
however, that one isolated set~(\ref{eq:CA}) of field values was singled out as a solution to the 
equations of motion, the level truncation approximation generically lifts a flat direction 
to an approximate one. In fact, the effective potential for the massless marginal deformation parameter 
in string field theory was studied in~\cite{SZ2,IqNaq2} and the authors concluded that the potential was 
not exactly flat to any given finite order in the level truncation scheme, though it became flatter 
as the level of approximation was increased. Therefore the kink-antikink configuration, when the distance 
between two branes is changed arbitrarily, 
will not solve the equations of motion. Nevertheless, we can `construct' 
such configurations by turning on the $\tau_2$ mode by hand. That is, we fix the value of $\tau_2$, 
say $\tau_2=0.1$, and then extremize the action with respect to the remaining 23 variables. 
Some of the resulting configurations are plotted in Figure~\ref{fig:move}. 
Though these configurations with $\tau_2\neq 0$ are not solutions, we can calculate their tensions 
anyway. The ratio $r$ of the tension of the kink-antikink to twice the tension of the BPS D$(p-1)$-brane, 
together with the location $x_0$ of the antikink defined to be a zero of the tachyon profile $t(x)$ with 
$t^{\prime}(x_0)<0$, is shown in Table~\ref{tab:B} as functions of the deformation parameter $\tau_2$. 
\begin{table}[htbp]
	\begin{center}
	\begin{tabular}{|r|c|c|}
	\hline 
	$\tau_2$ & $r$ (tension) & $x_0$ (zero of $t(x)$) \\
	\hline  \hline 
	0    & 0.987919 & 6.283 \\ \hline 
	0.01 & 0.987768 & 6.144 \\ \hline 
	0.02 & 0.987306 & 6.003 \\ \hline 
	0.03 & 0.986504 & 5.859 \\ \hline 
	0.04 & 0.985314 & 5.710 \\ \hline 
	0.05 & 0.983531 & 5.555 \\ \hline 
	0.06 & 0.981496 & 5.392 \\ \hline 
	0.07 & 0.978685 & 5.219 \\ \hline 
	0.08 & 0.975120 & 5.034 \\ \hline 
	0.09 & 0.970652 & 4.831 \\ \hline 
	0.10 & 0.964936 & 4.603 \\ \hline 
	0.11 & 0.958044 & 4.325 \\ \hline 
	$\ge 0.12$ & no solutions & no solutions \\ \hline
	\end{tabular}
	\end{center}
	\caption{We show the value of the ratio $r$ of the tension of the kink-antikink to twice the 
	BPS D$(p-1)$-brane tension, and the location $x_0$ of the antikink as functions of the 
	deformation parameter $\tau_2$. We found no solutions once $\tau_2$ reached or went beyond 0.12.}
	\label{tab:B}
\end{table}
Notice that $2 x_0$ is in fact equal to the distance between the kink and the antikink due to the 
symmetry $t(x)=t(-x)$. For the same reason, a configuration for a negative value of $\tau_2$ is 
equivalent to that for $|\tau_2|$, so we wrote the results only for $\tau_2\ge 0$ in Table~\ref{tab:B}. 
Our numerical algorithm ceased to converge when we chose the value equal to or more than 0.12 for $\tau_2$. 
From this table, one can find that the tension of the kink-antikink system gets \textit{lowered} as 
the kink and the antikink come closer. In order to find out more detailed relation between the 
tension $r$ and the distance $x_0$, we have plotted $\log_{10}(2\pi -x_0)$ versus $\log_{10}(0.987919-r)$ 
in Figure~\ref{fig:fitting} and fitted these data with a straight line. The result is 
\begin{equation}
\log_{10}(0.987919-r)=-2.09383+2.01593 \log_{10}(2\pi -x_0) \label{eq:CB}
\end{equation}
or
\begin{equation}
r=0.987919-0.00805649\times(2\pi -x_0)^{2.01593}. \label{eq:CC}
\end{equation}
As is seen from Figure~\ref{fig:fitting}, it nicely interpolates the data. We can read from the 
relation~(\ref{eq:CC}) that the tension of the system almost quadratically depends on the `co-distance' 
$2\pi -x_0$ between the kink and the antikink. Though we do not understand the precise meaning of 
the exponent `2' appearing in eq.(\ref{eq:CC}), it might have some physical meaning since 
the data points are well approximated by a straight line. One may possibly think 
that it is a manifestation of the attractive force between a brane and an antibrane. However, we 
do not think this to be the case. In addition to our claim made at the beginning of this 
sub-subsection, if the lifting of the potential originated from 
the attractive force, the $x_0$-dependence~(\ref{eq:CC}) of the potential would depend on the 
dimensionality $(p-1)$ of the branes which interact via the exchange of closed strings. The fact 
that the formula~(\ref{eq:CC}) does not contain the dimensionality at all implies that the lifting 
of the potential has nothing to do with the attractive force carried by the closed strings and is 
presumably an artifact of our level truncation approximation. In any case, 
what we have learned from the above analysis are:
\begin{itemize}
	\item A kink and an antikink are automatically placed at diametrically opposite points 
	of the circle if we solve the full set of equations of motion. 
	\item By setting $\tau_2$ to some fixed values and solving the equations of motion with 
	respect to the other variables, we can deform the above kink-antikink solution along the 
	quasi-flat direction of the potential representing the relative motion of the branes. 
	\item It seems that the tension of the kink-antikink system is lowered 
	as the kink approaches the antikink. 
\end{itemize}

\section{Summary}\label{sec:conclusion}
In this paper we have constructed a kink solution and a kink-antikink solution by applying the 
level truncation method to the Berkovits' version of open superstring field theory. The tensions of 
the solutions have agreed with those of the expected BPS D-branes with a remarkable precision (95\% 
and 99\% respectively). We have also tried to see the quasi-flat direction of the potential 
corresponding to the relative motion of the kink and the antikink, with some questions unanswered. 

We have focused our attention only on the classical configurations of the string field 
on a non-BPS D-brane. It would be interesting to study the fluctuation spectra around these 
solutions, as well as to construct a vortex solution after the tachyon condensation on a 
coincident brane-antibrane pair. 

\section*{Acknowledgements}
I am grateful to Prof. Tohru Eguchi for his insightful comments 
and careful reading of the manuscript. I would also like to thank 
Teruhiko Kawano, Seiji Terashima, Tadashi Takayanagi, Tadaoki Uesugi, 
Kent Ichikawa and Kazuhiro Sakai for valuable discussions.

\newpage

\section*{Appendices}
\renewcommand{\thesection}{\Alph{section}}
\setcounter{section}{0}

\section{Transformation of Vertex Operators}\label{sec:appA}
\subsection{Conformal transformations}
\begin{eqnarray}
g^{(n)}_k\circ(\xi ce^{-\phi}e^{ipX})(0)&=&\Bigg|\left(\frac{4}{n}\right)^{\ap p^2-\frac{1}{2}}
\Bigg| e^{-\pi i\left(\frac{k-1}{n}+\frac{1}{4}\right)}\xi ce^{-\phi}e^{ipX}\Bigl(g^{(n)}_k(0)\Bigr), 
\label{eq:AA} \\ g^{(n)}_k\circ(\xi\eta e^{ipX})(0)&=&\Bigg|\left(\frac{4}{n}\right)^{\ap p^2+1}
\Bigg| e^{2\pi i\left(\frac{k-1}{n}+\frac{1}{4}\right)}\left(\xi\eta-\frac{g^{(n)\prime\prime}_k(0)}
{2(g^{(n)\prime}_k(0))^2}\right)e^{ipX}\Bigl(g^{(n)}_k(0)\Bigr) \nonumber \\ &=& \Bigg|\left(
\frac{4}{n}\right)^{\ap p^2+1}\Bigg| e^{2\pi i\left(\frac{k-1}{n}+\frac{1}{4}\right)} \nonumber \\
& &\times\left(\xi\eta
-\frac{1}{2}e^{-2\pi i\frac{k-1}{n}}\right)e^{ipX}\Bigl(g^{(n)}_k(0)\Bigr), \label{eq:AB} \\
g^{(n)}_k\circ(\xi\partial\xi c\partial^2 ce^{-2\phi}e^{ipX})(0)&=&\Bigg|\left(\frac{4}{n}
\right)^{\ap p^2+1}\Bigg| e^{2\pi i\left(\frac{k-1}{n}+\frac{1}{4}\right)} \nonumber \\ & &\times 
\xi\partial\xi \left(c\partial^2 c-e^{-2\pi i\frac{k-1}{n}}c\partial c\right)e^{-2\phi}e^{ipX}
\Bigl(g^{(n)}_k(0)\Bigr), \label{eq:AC} \\ g^{(n)}_k\circ(\xi c(\partial e^{-\phi})\psi^Xe^{ipX})(0)
&=&\Bigg|\left(\frac{4}{n}\right)^{\ap p^2+1}\Bigg| e^{2\pi i\left(\frac{k-1}{n}+\frac{1}{4}\right)}
\nonumber \\ & &\times\xi c\left(\partial e^{-\phi}+\frac{1}{2}e^{-2\pi i\frac{k-1}{n}}e^{-\phi}
\right)\psi^Xe^{ipX}\Bigl(g^{(n)}_k\Bigr), \label{eq:AD} \\ 
g^{(n)}_k\circ(\xi ce^{-\phi}G^{\cM}e^{ipX})(0)&=&\Bigg|\left(\frac{4}{n}\right)^{\ap p^2+1}\Bigg| 
e^{2\pi i\left(\frac{k-1}{n}+\frac{1}{4}\right)}\xi ce^{-\phi}G^{\cM}e^{ipX}\Bigl(g^{(n)}_k(0)\Bigr), 
\label{eq:AE} \\ g^{(n)}_k\circ(\xi ce^{-\phi}\partial\psi^Xe^{ipX})(0)&=&\Bigg|\left(\frac{4}{n}
\right)^{\ap p^2+1}\Bigg|e^{2\pi i\left(\frac{k-1}{n}+\frac{1}{4}\right)} \nonumber \\ 
& &\times \xi ce^{-\phi}\left(\partial\psi^X+\frac{1}{2}e^{-2\pi i\frac{k-1}{n}}\psi^X
\right)e^{ipX}\Bigl(g^{(n)}_k(0)\Bigr), \label{eq:AF} \\ 
g^{(n)}_k\circ(\xi ce^{-\phi}\psi^X\partial Xe^{ipX})(0)&=&\Bigg|\left(\frac{4}{n}\right)^{\ap p^2
+1}\Bigg| e^{2\pi i\left(\frac{k-1}{n}+\frac{1}{4}\right)} \nonumber \\ 
& &\times\xi ce^{-\phi}\psi^X\left(\partial X-i\ap pe^{-2\pi i\frac{k-1}{n}}\right)e^{ipX}
\Bigl(g^{(n)}_k(0)\Bigr). \label{eq:AG}
\end{eqnarray}
In the above lines, we used (\ref{eq:M1}), (\ref{eq:M2}) so that 
\[ \frac{g^{(n)\prime\prime}_k(0)}{(g^{(n)\prime}_k(0))^2}=e^{-2\pi i\frac{k-1}{n}}. \]

\subsection{BRST transformations}
\begin{eqnarray}
Q_0(\xi ce^{-\phi}e^{ipX})&=&\left(\ap p^2-\frac{1}{2}\right)\xi c\partial ce^{-\phi}e^{ipX}, 
\label{eq:AH} \\ Q_1(\xi ce^{-\phi}e^{ipX})&=&-\sqrt{2\ap}p\ c\psi^Xe^{ipX}, \label{eq:AI} \\ 
Q_2(\xi ce^{-\phi}e^{ipX})&=&-\eta e^{\phi}e^{ipX}, \label{eq:AJ} \\
Q_0(\xi\eta e^{ipX})&=&\ap p^2\partial c\xi\eta e^{ipX}+\partial(c\xi\eta e^{ipX})-\frac{1}{2}
\partial^2c\ e^{ipX}, \label{eq:AK} \\
Q_1(\xi\eta e^{ipX})&=&\eta e^{\phi}G^{\mathrm{m}}e^{ipX}+\sqrt{2\ap}p\ \eta \partial 
(e^{\phi}\psi^X)e^{ipX}, \label{eq:AL} \\
Q_2(\xi\eta e^{ipX})&=&-2\eta\partial\eta e^{2\phi}be^{ipX}, \label{eq:AM} \\
Q_1(\partial\xi c\partial^2ce^{-2\phi}e^{ipX})&=&\sqrt{2\ap}p\ c\partial^2ce^{-\phi}\psi^Xe^{ipX}, \\
Q_2(\partial\xi c\partial^2ce^{-2\phi}e^{ipX})&=&(6\partial\eta+8\partial\phi\ \eta)ce^{ipX}, \\
Q_0(c\partial e^{-\phi}\psi^Xe^{ipX})&=&(\ap p^2+1)\partial cc\partial e^{-\phi}\psi^Xe^{ipX}
+\frac{1}{2}\partial^2cce^{-\phi}\psi^Xe^{ipX}, \\
Q_2\left(\xi c(\partial e^{-\phi})\psi^Xe^{ipX}\right)&=&(5\eta\partial e^{\phi}+4\partial\eta e^{\phi}
+2\eta bce^{\phi})\psi^Xe^{ipX}, \\
Q_2(\xi ce^{-\phi}G^{\cM}e^{ipX})&=&e^{\phi}\eta G^{\cM}e^{ipX}, \\
Q_2(\xi ce^{-\phi}(\partial\psi^X)e^{ipX})&=&e^{\phi}\eta(\partial\psi^X)e^{ipX}, \\
Q_2(\xi ce^{-\phi}\psi^X\partial Xe^{ipX})&=&e^{\phi}\eta\psi^X\partial Xe^{ipX}.
\end{eqnarray}

\section{Sample Calculations}\label{sec:appB}
In this section we exhibit some computations of the action. Firstly, at the level of approximation 
used in this paper, it is sufficient to keep only those terms with up to six string fields in 
the action~(\ref{eq:A}) due to the conservation of $\phi$-charge and the level truncation~\cite{BSZ}. 
We can further simplify the action by appealing to the twist symmetry and 
cyclicity properties~(\ref{eq:Q}), the result being 
\begin{eqnarray}
\hspace{-0.5cm}
S&=&\frac{1}{2g_o^2}\bllk\frac{1}{2}(\widehat{Q}_B
\widehat{\Phi})(\widehat{\eta}_0\widehat{\Phi})+\frac{1}{3}(\widehat{Q}_B
\widehat{\Phi})\widehat{\Phi}(\widehat{\eta}_0\widehat{\Phi})+\frac{1}{12}
(\widehat{Q}_B\widehat{\Phi})\left(\widehat{\Phi}^2(\widehat{\eta}_0
\widehat{\Phi})-\widehat{\Phi}(\widehat{\eta}_0\widehat{\Phi})\widehat{\Phi}
\right) \nonumber \\ & &{}+\frac{1}{60}(\widehat{Q}_B\widehat{\Phi})\left(
\widehat{\Phi}^3(\widehat{\eta}_0\widehat{\Phi})-3\widehat{\Phi}^2(
\widehat{\eta}_0\widehat{\Phi})\widehat{\Phi}\right) \nonumber \\ & &{}+
\frac{1}{360}(\widehat{Q}_B\widehat{\Phi})\left(\widehat{\Phi}^4(
\widehat{\eta}_0\widehat{\Phi})-4\widehat{\Phi}^3(\widehat{\eta}_0
\widehat{\Phi})\widehat{\Phi}+3\widehat{\Phi}^2(\widehat{\eta}_0\widehat{\Phi}
)\widehat{\Phi}^2\right)\brrk .
\end{eqnarray}
At the purely tachyonic level, $\widehat{\Phi}=T\otimes\sigma_1$, the nonvanishing terms are 
\begin{eqnarray*}
2g_o^2S_T&=&\frac{1}{2}\bllk\left(\widehat{Q}_B\widehat{\Phi}\right)\left(\widehat{\eta}_0
\widehat{\Phi}\right)\brrk+\frac{1}{12}\bllk\left(\widehat{Q}_B\widehat{\Phi}\right)\widehat{\Phi}^2
\left(\widehat{\eta}_0\widehat{\Phi}\right)-\left(\widehat{Q}_B\widehat{\Phi}\right)\widehat{\Phi}
\left(\widehat{\eta}_0\widehat{\Phi}\right)\widehat{\Phi}\brrk \\ 
&=&-\left\langle g^{(2)}_1\circ(Q_0T)\ g^{(2)}_2\circ(\eta_0T)\right\rangle-\frac{1}{6}\left\langle
g^{(4)}_1\circ(Q_2T)\ g^{(4)}_2\circ T\ g^{(4)}_3\circ T\ g^{(4)}_4\circ(\eta_0T)\right\rangle \\
& &{}-\frac{1}{6}\left\langle g^{(4)}_1\circ(Q_2T)\ g^{(4)}_2\circ T\ g^{(4)}_3\circ(\eta_0 T)\ 
g^{(4)}_4\circ T \right\rangle , 
\end{eqnarray*}
where we have taken the trace and used the decomposition~(\ref{eq:B}) of the BRST charge. Let us 
show how to calculate them. Since the BRST charge $Q_B$ and $\eta_0$ commute with the Virasoro 
generators $L_n^{\mathrm{tot}}$, we can conformally transform $T$ in advance, namely 
\[ g^{(n)}_k\circ(Q_BT)=Q_B\left(g^{(n)}_k\circ T\right), \quad g^{(n)}_k\circ (\eta_0T)=\eta_0
\left(g_k^{(n)}\circ T\right). \]
Using the definition~(\ref{eq:Z}) of $T$ and the partial integration~(\ref{eq:R}), the 
2-point correlator becomes 
\begin{eqnarray}
-\left\langle g^{(2)}_1\circ (Q_0T)\ g^{(2)}_2\circ (\eta_0T)\right\rangle&=&-\sum_{m,n}t_mt_n
\Bigg|\left(\frac{4}{2}\right)^{\frac{\ap}{R^2}(m^2+n^2)-1}\Bigg|e^{-\pi i\left(\frac{1}{4}+\frac{1}{2}
+\frac{1}{4}\right)} \label{eq:CI} \\
& &\times\left\langle\xi ce^{-\phi}e^{i\frac{m}{R}X}(z_1)Q_0\left(
ce^{-\phi}e^{i\frac{n}{R}X}\right)(z_2)
\right\rangle , \nonumber
\end{eqnarray}
where we have used the conformal transformation~(\ref{eq:AA}) for $n=2$, and defined 
\[ z_i=g^{(2)}_i(0)=\left\{
	\begin{array}{ccc}
	1 & \mathrm{for} & i=1 \\
	-1 & \mathrm{for} & i=2
	\end{array}
\right. .\]
The correlator in (\ref{eq:CI}) is calculated as 
\begin{eqnarray*}
& &\left(\frac{\ap}{R^2}n^2-\frac{1}{2}\right)\left\langle\xi(z_1)c(z_1)\partial cc(z_2)
e^{-\phi(z_1)}e^{-\phi(z_2)}e^{i\frac{m}{R}X(z_1)}e^{i\frac{n}{R}X(z_2)}\right\rangle \\
&=&\left(\frac{\ap}{R^2}n^2-\frac{1}{2}\right)\frac{(z_1-z_2)^2}{2}\frac{1}{z_1-z_2}\langle
\xi(z_1)\partial^2c\partial cc(z_2)e^{-2\phi(z_2)}\rangle_{\mathrm{ghost}} \\
& &\times |z_1-z_2|^{\frac{2\ap}{R^2}mn}\ 2\pi RV_p\delta_{m+n,0} \\
&=&-2\pi RV_p\delta_{m+n,0}\cdot 2^{-\frac{2\ap}{R^2}n^2+1}\left(\frac{\ap}{R^2}n^2-\frac{1}{2}\right),
\end{eqnarray*}
where we have used the action~(\ref{eq:AH}) of $Q_0$ on $T$, the OPEs~(\ref{eq:E6}), (\ref{eq:E1}), 
the normalization~(\ref{eq:P}) and 
\[ \langle e^{iqX}\rangle_{\mathrm{matter}}=(2\pi)^{p+1}\delta^{p+1}(q) \]
on a D$p$-brane. Multiplying it by the prefactors of~(\ref{eq:CI}), we finally obtain the expression 
for the quadratic term in $t$ as 
\begin{equation}
2g_o^2S_T^{(2)}=2\pi RV_p\sum_n\left(-\frac{\ap}{R^2}n^2+\frac{1}{2}\right)t_{-n}t_n. \label{eq:CG}
\end{equation}
Now we turn to the quartic terms. We need to calculate 
\begin{eqnarray*}
\cA(a,b)&=&\langle g^{(4)}_1\circ (Q_2T)\ g^{(4)}_2\circ T\ g^{(4)}_a\circ T\
g^{(4)}_b\circ (\eta_0T)\rangle \\
&=&\sum t_{n_1}t_{n_2}t_{n_a}t_{n_b}\Bigg|\left(\frac{4}{4}\right)^{\frac{\ap}{R^2}(n_1^2+n_2^2+
n_a^2+n_b^2)-2}\Bigg|e^{-\pi i\frac{1+2+3+4}{4}} \\ & &\times \left\langle Q_2\left(\xi ce^{-\phi}
e^{i\frac{n_1}{R}X}\right)(z_1)\xi ce^{-\phi}e^{i\frac{n_2}{R}X}(z_2)\xi ce^{-\phi}e^{i\frac{n_a}{R}X}(z_a)
ce^{-\phi}e^{i\frac{n_b}{R}X}(z_b)\right\rangle \\ 
&=&\sum t_{n_1}t_{n_2}t_{n_a}t_{n_b}e^{-\frac{\pi i}{2}}
\left\langle(-\eta e^{\phi})(z_1)\xi ce^{-\phi}(z_2)\xi ce^{-\phi}(z_a)ce^{-\phi}(z_b)
\right\rangle_{\mathrm{g}} \\ & &\times\left\langle e^{i\frac{n_1}{R}X_1}e^{i\frac{n_2}{R}X_2}
e^{i\frac{n_a}{R}X_a}e^{i\frac{n_b}{R}X_b}\right\rangle_{\mathrm{m}} \\ 
&=&-i\sum 
t_{n_1}t_{n_2}t_{n_a}t_{n_b}\langle\eta_1\xi_2\xi_a(-1)c_2c_ac_be^{\phi_1}e^{-\phi_2}e^{-\phi_a}
e^{-\phi_b}\rangle_{\mathrm{g}} \\ & &\times 2\pi RV_p\delta_{n_1+n_2+n_a+n_b,0}
|z_{12}|^{\frac{2\ap}{R^2}n_1n_2}|z_{1a}|^{\frac{2\ap}{R^2}n_1n_a}|z_{1b}|^{\frac{2\ap}{R^2}n_1n_b}
|z_{2a}|^{\frac{2\ap}{R^2}n_2n_a}|z_{2b}|^{\frac{2\ap}{R^2}n_2n_b}|z_{ab}|^{\frac{2\ap}{R^2}n_an_b},
\end{eqnarray*}
where we have defined 
\[ z_j=g^{(4)}_j(0)=\left\{
	\begin{array}{ccc}
	1  & \mathrm{for} & j=1 \\
	i  & \mathrm{for} & j=2 \\
	-1 & \mathrm{for} & j=3 \\
	-i & \mathrm{for} & j=4
	\end{array}
\right. , \]
and $z_{jk}=z_j-z_k$. The subscript of the field denotes its position: \textit{e.g.} $\xi_a=
\xi(z_a)$. The ghost correlator can be evaluated as 
\begin{eqnarray*}
&-&\left\langle\left(\frac{\xi_a}{z_{12}}-\frac{\xi_2}{z_{1a}}\right)\cdot z_{2a}z_{2b}z_{ab}\frac{1}{2}
\partial^2c\partial cc\cdot \frac{z_{12}z_{1a}z_{1b}}{z_{2a}z_{2b}z_{ab}}e^{-2\phi}
\right\rangle_{\mathrm{g}} \\ &=&z_{2a}z_{1b}
\end{eqnarray*}
using the relevant OPEs. Then we find 
\begin{eqnarray*}
\cA(3,4)&=&2\times 2\pi RV_p\sum\delta_{n_1+n_2+n_3+n_4}t_{n_1}t_{n_2}t_{n_3}t_{n_4}\cdot
2^{\frac{\ap}{R^2}(n_1n_2+n_2n_3+n_3n_4+n_4n_1+2n_1n_3+2n_2n_4)}, \\
\cA(4,3)&=&4\times 2\pi RV_p\sum\delta_{n_1+n_2+n_3+n_4}t_{n_1}t_{n_2}t_{n_3}t_{n_4}\cdot
2^{\frac{\ap}{R^2}(n_1n_2+n_2n_3+n_3n_4+n_4n_1+2n_1n_3+2n_2n_4)}.
\end{eqnarray*}
The quartic terms in the action can be written as 
\begin{eqnarray}
2g_o^2S_T^{(4)}&=&-\frac{1}{6}\cA(3,4)-\frac{1}{6}\cA(4,3) \nonumber \\
&=&-2\pi RV_p\sum\delta_{n_1+n_2+n_3+n_4}t_{n_1}t_{n_2}t_{n_3}t_{n_4}\cdot 2^{\frac{\ap}{R^2}
((n_1+n_3)^2-n_1^2-n_2^2-n_3^2-n_4^2)}, \label{eq:CH}
\end{eqnarray}
where the exponent was put to the above form using the momentum conservation. Combining eq.(\ref{eq:CG}) 
and (\ref{eq:CH}), we finally get the purely tachyonic string field theory action~(\ref{eq:BH}). 
\medskip

Of course, we must also calculate a large number of correlators which contain non-tachyonic modes, 
even though only those terms with $-2$ units of total $\phi$-charge will survive. As an example, 
we will illustrate the calculations with the following quartic coupling 
\[ \cB(a,b,c,d)=\left\langle g^{(4)}_a\circ(Q_1T)\ g^{(4)}_b\circ T\ g^{(4)}_c\circ E\ 
g^{(4)}_d\circ(\eta_0J)\right\rangle. \]
Using the conformal transformation laws~(\ref{eq:AA}),(\ref{eq:AB}),(\ref{eq:AG}) and the 
action~(\ref{eq:AI}) of $Q_1$ on $T$, it can be calculated as 
\begin{eqnarray*}
\cB(a,b,c,d)&=&\frac{i}{\sqrt{\ap}}\sum t_{n_a}t_{n_b}e_{n_c}j_{n_d}\Bigg|\left(\frac{4}{4}
\right)^{\frac{\ap}{R^2}(n_a^2+n_b^2+n_c^2+n_d^2)+1}\Bigg|e^{\frac{2\pi i}{4}\left(-\frac{a}{2}
-\frac{b}{2}+c+d\right)} \\ & &\times\Biggl\langle Q_1\left(\xi ce^{-\phi}e^{i\frac{n_a}{R}X}
\right)(z_a)\xi ce^{-\phi}e^{i\frac{n_b}{R}X}(z_b)\left(\xi\eta(z_c)-\frac{1}{2}e^{\frac{\pi i}{2}
(1-c)}\right) \\ & &\ \ \times e^{i\frac{n_c}{R}X(z_c)}ce^{-\phi}\psi^X\left(\partial X-i\frac{\ap}{R}
n_de^{\frac{\pi i}{2}(1-d)}\right)e^{i\frac{n_d}{R}X}(z_d)\Biggr\rangle \\
&=&\frac{i}{\sqrt{\ap}}\sum t_{n_a}t_{n_b}e_{n_c}j_{n_d}e^{\frac{\pi i}{2}\left(-5+\frac{3}{2}(c+d)\right)}
\left(-\frac{\sqrt{2\ap}}{R}n_a\right) \\ & &\times\left\langle c\psi^X(z_a)\xi ce^{-\phi}(z_b)
\left(\xi\eta(z_c)-\frac{1}{2}e^{\frac{\pi i}{2}(1-c)}\right)ce^{-\phi}\psi^X(z_d)
\right\rangle_{\mathrm{ghost},\psi} \\ & &\times\left\langle\left(\partial X(z_d)-i\frac{\ap}{R}n_d
e^{\frac{\pi i}{2}(1-d)}\right)e^{i\frac{n_a}{R}X_a}e^{i\frac{n_b}{R}X_b}e^{i\frac{n_c}{R}X_c}
e^{i\frac{n_d}{R}X_d}\right\rangle_X, 
\end{eqnarray*}
where we have arranged the phase factor in the above way using the fact $a+b+c+d=10$. ($\{a,b,c,d\}$ is 
a permutation of $\{1,2,3,4\}$.) Taking care not to forget the minus signs when fermionic fields pass 
each other, the ghost- and $\psi$-part becomes 
\begin{eqnarray*}
&-&\left\langle\xi_b\left( :\xi_c\eta_c:-\frac{1}{2}e^{\frac{\pi i}{2}(1-c)}\right)c_ac_bc_d
e^{-\phi_b}e^{-\phi_d}\psi^X_a\psi^X_d\right\rangle \\ &=&-\left\langle\left(-\frac{\xi_c}{z_{bc}}
-\frac{1}{2}e^{\frac{\pi i}{2}(1-c)}\xi_b\right)\cdot z_{ab}z_{ad}z_{bd}\frac{1}{2}\partial^2c
\partial cc\cdot\frac{1}{z_{bd}}e^{-2\phi}\cdot \frac{1}{z_{ad}}\right\rangle \\
&=&-\left(\frac{1}{z_{bc}}+\frac{1}{2}e^{\frac{\pi i}{2}(1-c)}\right)z_{ab}. 
\end{eqnarray*}
For the $X$-part, the OPE~(\ref{eq:E2}) gives 
\begin{eqnarray*}
& &\left\{\frac{i}{R}(-2\ap)\left(\frac{n_a}{z_{da}}+\frac{n_b}{z_{db}}+\frac{n_c}{z_{dc}}\right)-i
\frac{\ap}{R}n_de^{\frac{\pi i}{2}(1-d)}\right\}2\pi RV_p\delta_{n_a+n_b+n_c+n_d} \\
& &\times |z_{ab}|^{\frac{2\ap}{R^2}n_an_b}p|z_{ac}|^{\frac{2\ap}{R^2}n_an_c}
|z_{ad}|^{\frac{2\ap}{R^2}n_an_d}
|z_{bc}|^{\frac{2\ap}{R^2}n_bn_c}|z_{bd}|^{\frac{2\ap}{R^2}n_bn_d}|z_{cd}|^{\frac{2\ap}{R^2}n_cn_d} \\
&=&-i\frac{\ap}{R}\left(\frac{2n_a}{z_{da}}+\frac{2n_b}{z_{db}}+\frac{2n_c}{z_{dc}}+n_de^{\frac{\pi i}{2}
(1-d)}\right)2\pi RV_p\delta_{\Sigma n}\ 2^{\frac{\ap}{R^2}\left\{(n_1+n_3)^2-\Sigma n^2\right\}}.
\end{eqnarray*}
Collecting these factors, we finally obtain 
\begin{eqnarray*}
\cB(a,b,c,d)&=&2\pi RV_p\frac{\sqrt{2}\ap}{R^2}z_{ab}\left(\frac{1}{z_{bc}}+\frac{1}{2}
e^{\frac{\pi i}{2}(1-c)}\right)e^{\frac{\pi i}{2}\left(-5+\frac{3}{2}(c+d)\right)}\sum_{\{n\}}
\delta_{\Sigma n}\ 2^{\frac{\ap}{R^2}\left\{(n_1+n_3)^2-\Sigma n^2\right\}} \\ & &\times
n_at_{n_a}t_{n_b}e_{n_c}j_{n_d}\left(\frac{2n_a}{z_{da}}+\frac{2n_b}{z_{db}}+\frac{2n_c}{z_{dc}}
+n_de^{\frac{\pi i}{2}(1-d)}\right).
\end{eqnarray*}
Repeating the computations this way, one could write down the component form of the action. 
Concerning our result, 
\renewcommand{\labelenumi}{(\theenumi)}
\begin{enumerate}
	\item our action reproduces the level $\left(\frac{3}{2},3\right)$ tachyon 
	potential found in~\cite{BSZ} when we restrict all fields in the action to zero 
	momentum $(n=0)$ sector, and
	\item our action is manifestly real if the component fields defined in~(\ref{eq:Z}) 
	are real. 
\end{enumerate}
These features suggest that our action is correct. In particular, the fact (1) 
strongly supports the correctness of our calculations in the ghost sector. 

\section{On the use of conservation laws}
In ref.\cite{RZ}, a convenient computational scheme, using the conservation laws, for string field 
theory has been developed. Although the authors of~\cite{RZ} made a brief comment on the 
applicability of the conservation laws to superstring field theory, we have judged that it is not 
very advantageous to use the conservation laws for calculating the string vertices 
in Berkovits' superstring field theory for the following reasons. 

The first point, though not so problematic, is that every string vertex contains the insertion of the 
BRST charge $Q_B$, as opposed to the cubic vertex in bosonic cubic string field theory~\cite{csft}. 
Since it takes the complicated form 
\begin{eqnarray*}
Q_B^{\mathrm{holom.}}&=&\sum_mc_{-m}L_m^{\mathrm{m}}+\sum_r\gamma_{-r}
G_r^{\mathrm{m}}-\sum_{m,n}\frac{n-m}{2}\ca b_{-m-n}c_mc_n\ca 
\\& &+\sum_{m,r}\left[\frac{2r-m}{2}\ca\beta_{-m-r}c_m\gamma_r\ca -\ca b_{-m}
\gamma_{m-r}\gamma_r\ca \right]+a^{\mathrm{g}}c_0
\end{eqnarray*}
in the operator representation, we want to avoid making explicit use of it. The second point is the 
problem of dealing with the states in the ``large" Hilbert space. If we were working in the 
``small" Hilbert space, it would be most convenient to write down the relevant conservation laws 
for the $\beta,\gamma$ oscillators. However, once we carry out the bosonization~(\ref{eq:D}) of 
the $\beta,\gamma$ ghosts in order to incorporate the zero mode of $\xi$, we will be confronted with 
the treatment of the factor $e^{\ell\phi}$. Since we are not given an oscillator form of it, we have to 
handle it in a similar way as the momentum factor $e^{ipX}$. We are not familiar with such a formulation. 

The third problem consists in the way of finding the conservation laws. In the case of bosonic 
cubic string field theory, the Virasoro conservation laws of the general form 
\begin{equation}
\langle V_3|L_{-k}^{(2)}=\langle V_3|\left(A^k\cdot c+\sum_{n\ge 0}a_n^kL_n^{(1)}+\sum_{n\ge 0}b_n^k
L_n^{(2)}+\sum_{n\ge 0}c_n^kL_n^{(3)}\right) \label{eq:CE}
\end{equation}
have been derived in~\cite{RZ} by introducing a holomorphic vector field $v(z)$. To obtain the Virasoro 
conservation law for the $k=2$ case, we had to choose the vector field to be 
\begin{equation}
v(z)=\frac{(z-z_1)(z-z_3)}{z-z_2}, \label{eq:CF}
\end{equation}
where $z_i$ denotes the insertion point of the $i$-th vertex operator on the boundary of the global disk. 
That is, $v(z)$ has zeroes at punctures 1 and 3, and has a simple pole at puncture 2. If we wish 
to extend the conservation law~(\ref{eq:CE}) for $k=2$ to, say, the 5-point vertex $\langle V_5|$, 
it is expected that the holomorphic vector field must be of the form 
\[ v(z)=\frac{(z-z_1)(z-z_3)(z-z_4)(z-z_5)}{z-z_2}. \]
However, it does not respect the holomorphicity at infinity, which is expressed as 
\[ \lim_{z\to\infty} z^{-2}v(z)<+\infty. \]
Hence it seems that it is impossible to write down Virasoro conservation laws with a small value of $k$ 
for the string vertices of sufficiently high order. 

%%%%%refs.

\end{document}